\documentclass[10pt,journal,final]{IEEEtran}
\usepackage{setspace,cite,graphicx,epstopdf,subfigure,threeparttable,booktabs,amsmath,amsfonts,amssymb,bm,booktabs}
\usepackage{hyperref}
\hypersetup{
  pdfauthor={...},
  pdftitle={...},
  pdfsubject={...},
  urlcolor=blue,
}
\usepackage{caption2}
\usepackage{color}
\usepackage{amssymb}
\usepackage{hyperref}
\usepackage{url}
\usepackage{epsfig}
\usepackage{graphicx}
\usepackage{amsmath}
\usepackage{amssymb}
\usepackage{amsfonts}
\usepackage{amsmath,graphicx,subfigure,threeparttable,booktabs,float,stfloats,multirow}
\usepackage{amssymb}
\usepackage{diagbox}
\usepackage{bm}
\usepackage{algorithm}
\usepackage{algorithmic}
\usepackage{multirow}
\usepackage{xcolor}

\usepackage{graphics}
\usepackage{indentfirst}
\usepackage{lineno}
\ifCLASSINFOpdf
\else
\fi
\begin{document}
\title{A Fusion Adversarial Underwater Image Enhancement Network with a Public Test Dataset}
\author{\IEEEauthorblockN{Hanyu Li, Jingjing Li, and Wei Wang}\\
\thanks{Corresponding author: *Hanyu Li (email:lihanyu1204@gmail.com).

H. Li is with Nanjing University of Information Science and Technology, Nanjing, China.

J. Li and W. Wang are engineers graduated from Nanjing University of Information Science and Technology, Nanjing, China.}

 }
\maketitle
\begin{abstract}
Underwater image enhancement algorithms have attracted much attention in underwater vision task.
    However, these algorithms are mainly evaluated on different data sets and different metrics.
In this paper, we set up an effective and public underwater test dataset named U45 including the color casts, low contrast and
haze-like effects of underwater degradation and propose a fusion adversarial  network for enhancing underwater images.
   Meanwhile, the well-designed the adversarial  loss including ${\cal L}_{gt}$ loss and ${\cal L}_{fe}$ loss is presented to focus on image features of ground truth, and image
features of the image enhanced by fusion enhance method, respectively.
  The proposed network corrects color casts effectively and owns faster testing time with fewer parameters.
Experiment results on U45 dataset demonstrate that the proposed method achieves better or comparable performance than the other state-of-the-art methods in terms of qualitative and quantitative evaluations.
  Moreover, an ablation study demonstrates the contributions of each component, and the application test further shows the effectiveness of the enhanced images.

\end{abstract}

\IEEEpeerreviewmaketitle
\section{\textbf{Introduction}}

\begin{figure}
\centering
\includegraphics[height=65mm,width=95mm,angle=0]{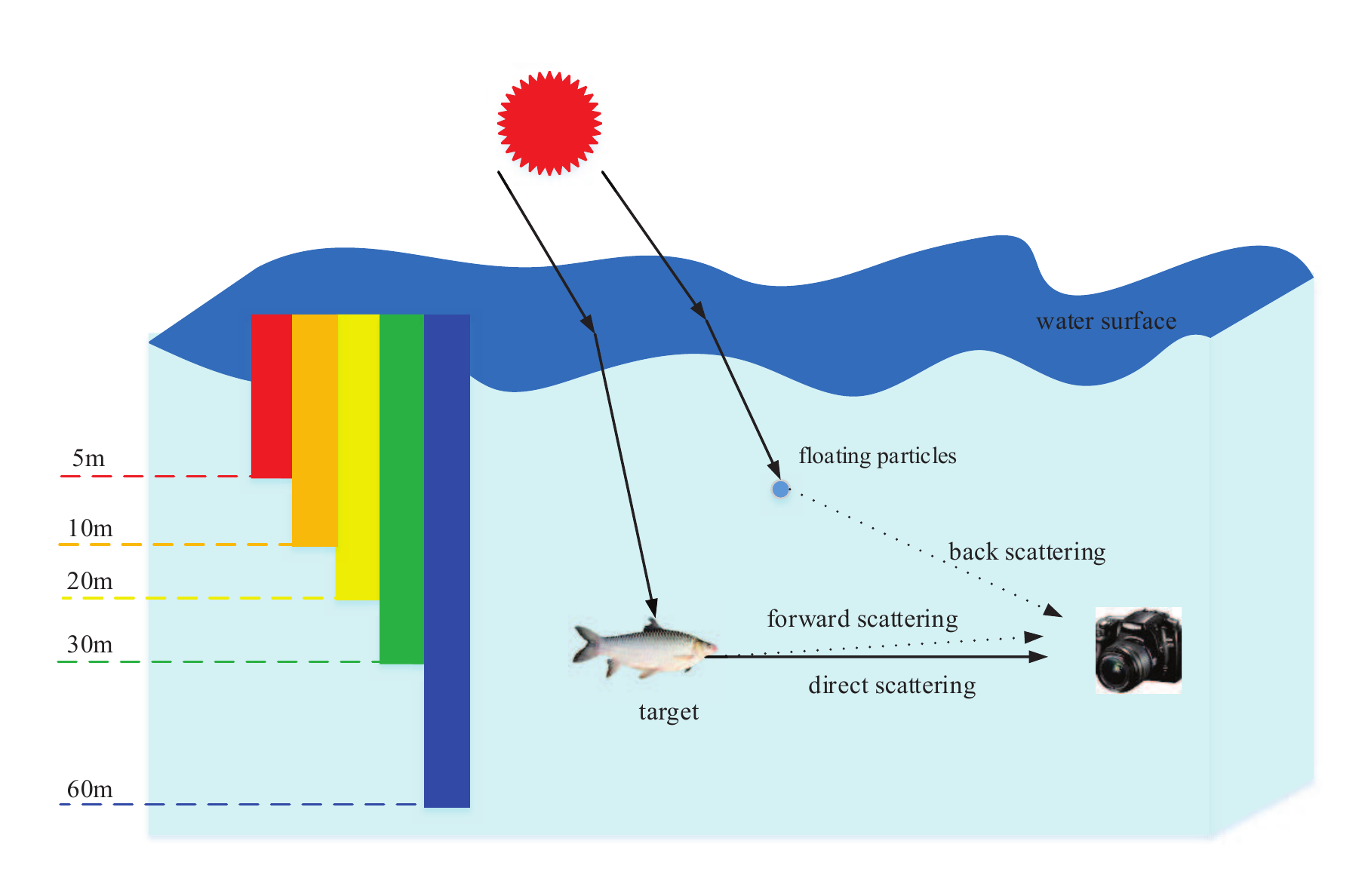}
\vspace{-6mm}
\caption{The simplified underwater optical imaging model.}
\label{fig1}
\end{figure}

The underwater optical imaging quality is of great significance to the exploration and utilization of deep ocean \cite{1}.
    However, raw underwater images seldom fulfill the requirements concerning low-level and high-level vision tasks because of the serious underwater degradation model depicted in Fig. \ref{fig1}.
The underwater images are rapidly degraded by two major factors.
    Due to the depth, light conditions, water type, and different light wavelengths, the color tones of underwater images are often distorted.
For example, in clean water, red light disappears first at 5 m water depth.
  As the depth of water increases, orange, yellow, and green lights disappear, respectively.
The blue light has the shortest wavelength and travels the furthest distance in the water \cite{2}.
    In addition, both suspended particles and water affect the scene contrast and produce haze-like effects by absorbing and scattering light to the camera lenses \cite{3}.

  To solve this problem, existing underwater image enhancement methods can be divided into three categories: non model-based methods \cite{4,5,6}, model-based methods \cite{2,7,8,9,10,11} and deep learning-based methods \cite{12,13,14,15,16,17,18,19}.
The non model-based methods focus on adjusting image pixel values to produce a subjectively and visually appealing image without modeling the underwater image formation process.
    Ancuti \textit{et al.} \cite{4} proposed a multi-scale fusion method by blending a white balanced version and a filtered version.
    The model-based methods recover underwater images by constructing the degradation model and then estimate model parameters from prior assumptions.
Peng \textit{et al.} \cite{10} presented an underwater depth estimation prior based on image blurriness and light absorption, which can be employed in the underwater optical imaging model to improve underwater image quality.
Due to the lack of abundant training data, the pixel values adjustment and physical priors can not perform well in various underwater scenes.
    Deep learning \cite{20} obtains convincing success on low-level vision tasks, such as image defogging \cite{21}, image deraining \cite{22}, and image deblurring \cite{23}, and some researchers apply deep learning to underwater image processing.
    Li \textit{et al.} \cite{12} proposed a weakly supervised underwater color transfer model based on cycle-consistent generative adversarial network (CycleGAN) \cite{24} and multi-term loss function.
Contrast to a previous application, our model blends two inputs to correct color effectively and owns faster testing time with fewer parameters.

\begin{figure*}[!htb]
\centering
\includegraphics[height=90mm,width=170mm,angle=0]{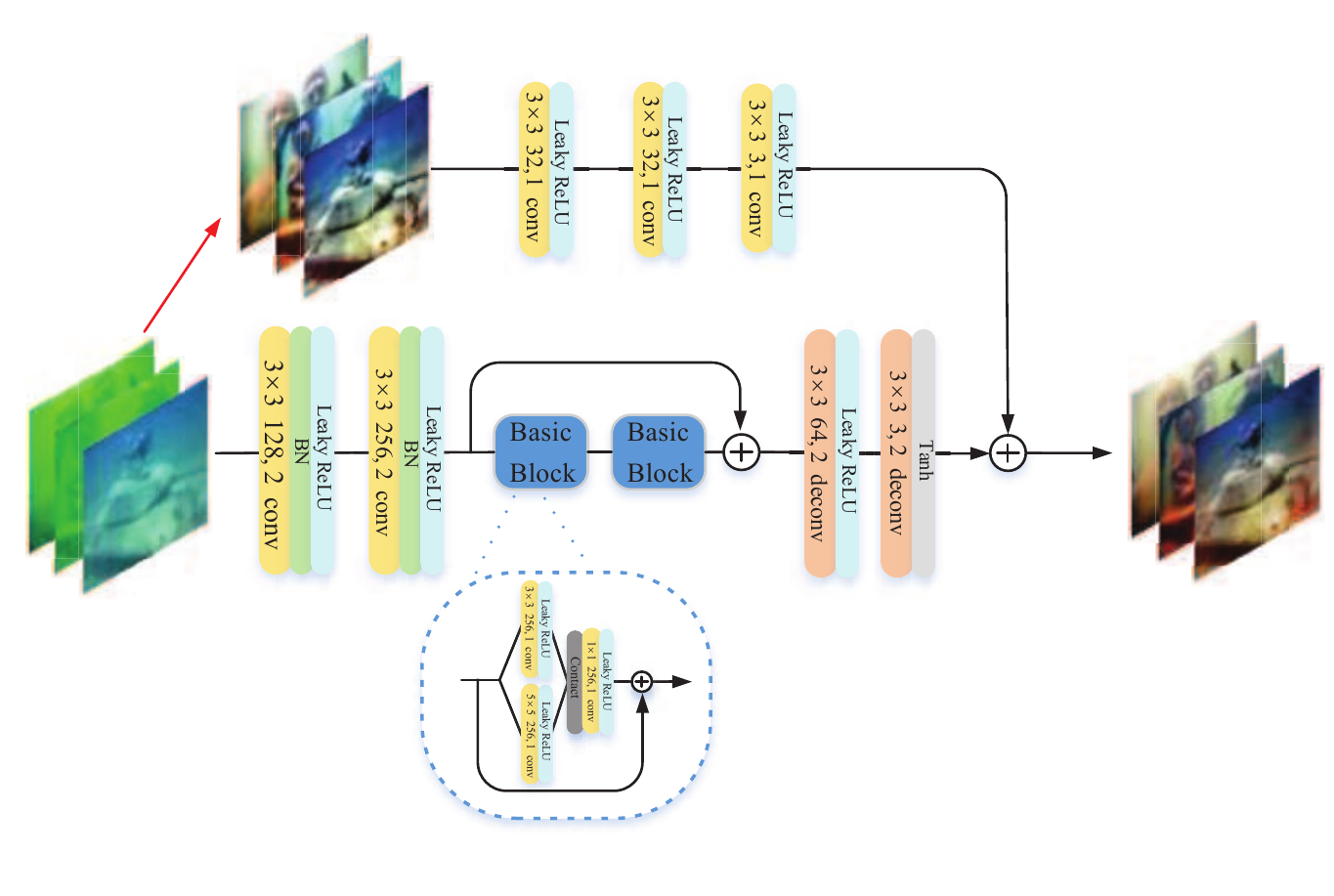}
\vspace{-3mm}
\caption{Architecture of generator network. ``conv'' denotes convolution layer while ``deconv'' denotes deconvolution layer.
Red arrow represents the process of fusion enhance method, and ``BN'' represents batch normalization.
 $3\times3$ denotes convolution kernel size, 128 denotes the number of convolution kernel, and 1 denotes convolution stride.}
\label{fig2}
\end{figure*}

In this paper, we propose a novel fusion generative adversarial  network (FGAN), and the main contributions of this paper are summarized as follows:

\begin{itemize}
\item  To the best of our knowledge, this is the first attempt to consider the multiple inputs fusion GAN in underwater image enhancement task.
  Besides, the simple and effective FGAN owns fast processing speed and fewer parameters without manually adjusting the image pixel values or designing the prior assumptions.

\item   The well-designed objective function is leveraged for preserving image content, and the spectral normalization is utilized to improve image quality. In addition, an ablation study shows the effect of each component in the proposed network.

\item   In order to evaluate quantitatively different algorithms, we set up an effective and public underwater test dataset (U45) including the color casts, low contrast and haze-like effects of underwater degradation.
   The enhanced images on U45 dataset and videos demonstrate the superiority of the proposed method in both qualitative and quantitative evaluations.
Furthermore, the enhanced results by the proposed method could benefit low-level and high-level vision tasks, such as canny edge detection and object detection.

\end{itemize}

\section{\textbf{Methodology}}
To solve color casts, low contrast and haze-like effects, we effectively blend multiple inputs and the generative adversarial  network \cite{25}.
  As shown in Fig. \ref{fig2}, the generator network employs two inputs in a fully convolutional network and combines two simple basic blocks. ${\cal L}_{gt}$ loss and ${\cal L}_{fe}$
  loss preserve image features of ground truth $x$, and preserve image features of enhanced images $x_{fe}$ produced by fusion enhance method \cite{4}, respectively.

\begin{table*}[!htb]
\small
\centering\
 \caption{Underwater image quality evaluation of different enhancement methods on U45 dataset.}
  \begin{tabular}{cccccccccccc}
    \toprule
    Dataset&Metric&Raws&FE&RB&UDCP&UIBLA&DPATN&CycleGAN&WSCT&UGAN&FGAN\\
    \midrule
    \multirow{5}*{\shortstack{Green}}&UCIQE  &0.5036 &{\color{red} 0.6444} &0.6029 &0.5896 &0.5732 &{\color{blue} 0.6369} &0.5972 &0.5717 &0.6039 &0.5935\\
    &UIQM  &1.4536    &3.4962   &  {\color{blue} 4.7391}  &2.8791    &1.9330    &4.0728    &4.0209    &2.4742    &4.7144    & {\color{red} 4.8362} \\
    &UICM  &-111.2208    &-46.5145    &-2.9019    &-71.3505    &-76.6133    &-53.4647    &-22.7221    &-82.3234    &-3.6953    &-0.0883   \\
    &UIConM &0.6789    &0.7422    &0.7600    &0.7976    &0.5733    &0.9792    &0.7285    &0.7504    &0.7557    &0.7693    \\
    &UISM  &7.3238    &7.2951    &7.1235    &6.9063    &6.9213    &7.0420    &6.9660    &7.1552    &7.1685    &7.0718  \\
    \midrule
    \multirow{5}*{\shortstack{Blue}}&UCIQE  &0.4905 &{\color{blue} 0.6568}  &0.6131 &0.6002 &0.5569 &{\color{red} 0.6693} &0.5749 &0.5663 &0.6151 &0.5885\\
    &UIQM  &2.5669    &4.2666    &4.9896    &4.3057    &2.8483    &4.3456    &4.5665    &3.3357    &  {\color{blue} 5.0442}   & {\color{red} 5.2583} \\
    &UICM  &-67.3196    &-30.7335    &-2.2822    &-25.1365    &-72.2416    &-18.4767    &-18.0013    &-62.9050    &-4.9056    &1.7935    \\
    &UIConM &0.6466    &0.8275    &0.8125    &0.8272   &0.7813    &0.7650    &0.8384    &0.8338    &0.8533    &0.8689    \\
    &UISM  &7.2921    &7.3641    &7.2776    &6.9665    &7.0846    &7.0846    &7.0321    &7.2082    &7.2186    &7.1148  \\
    \midrule
    \multirow{5}*{\shortstack{Haze-like}}&UCIQE  &0.4502 & {\color{blue} 0.6336}   &0.6022    &0.5719    &0.5668    & {\color{red} 0.6473}  &0.5785    &0.5844    &0.6141    &0.5925\\
    &UIQM  &3.0212    &4.5659    &5.0887    &4.4425    &4.0316    & {\color{red} 5.2218}   &4.6104    &4.0479    &5.1675    & {\color{blue} 5.2106}  \\
    &UICM  &-47.8476    &-21.6721    &-1.6179    &-22.6640    &-30.3830    &-1.0538    &-7.0986    &-32.7013    &7.3280    &8.7317    \\
    &UIConM &0.6176    &0.8509    &0.8409    &0.8197    &0.7807    &0.8658    &0.7674    &0.7984    &0.7949    &0.8042    \\
    &UISM  &7.3227    &7.2289    &7.2058    &7.2836    &7.1020    &7.3007    &6.9999    &7.1645    &7.1749    &7.0750  \\
    \midrule
    \multirow{5}*{\shortstack{Total}}&UCIQE  &0.4814    & {\color{blue} 0.6449}   &0.6061    &0.5872    &0.5656    & {\color{red} 0.6512}   &0.5835    &0.5741    &0.6110    &0.5915\\
    &UIQM  &2.3472    &4.1096    & 4.9391    &3.8758    &2.9376    &4.5467    &4.3993    &3.2859    &{\color{blue}4.9754 }   & {\color{red} 5.1017} \\
    &UICM  &-75.4627    &-32.9734    &-2.2673    &-39.7170    &-59.7460    &-24.3317    &-15.9406    &-59.3099    &-0.4243    &3.4790    \\
    &UIConM &0.6477    &0.8069    &0.8045    &0.8148    &0.7118    &0.8700    &0.7781    &0.7942    &0.8013    &0.8141    \\
    &UISM  &7.3129    &7.2960    &7.2023    &7.0521    &7.0359    &7.1871    &6.9993    &7.1760    &7.1874    &7.0872  \\
    \bottomrule
 \end{tabular}
 \label{tab0}
\end{table*}

\subsection{Network architecture}
  As discussed in the literature \cite{18}, fusion enhance method \cite{4} performs relatively well in most cases, thus we take the image enhanced by fusion enhance method as anther input of network.
     In this paper, the outputs of the simple network fed by $x_{fe}$ are added to the outputs of the network fed by raw underwater images, which uses less computer resources than the elementwise product of matrices employed by DUIENet \cite{18}.

Combined with inception architecture \cite{26} and shortcut connection \cite{27}, Fig. \ref{fig2} depicts the detailed structure of our basic block, which has different kernel sizes to detect the feature-maps at different scales. Each layer in block uses the convolutional kernel with stride 1 to facilitate the concatenation operation. Besides, the last $1\times1$ convolution of basic block reduces the output feature-maps to the number of input feature-maps of basic block, and thus improves computational efficiency and facilitates residual learning.
  The generator network employing two basic blocks achieves comparable performance without introducing too many blocks.

The discriminator network utilizes five convolutional layers with spectral normalization \cite{28}, similar to the work of $70\times70$ PatchGAN, where PatchGAN is first used in pix2pix \cite{29} and then extends to apply in later CycleGAN \cite{24}.
      Furthermore, spectral normalization is computationally light and easy to implement.
   In ablation study, we notice that the discriminator with spectral normalization has better objective quality score than the discriminator without spectral normalization.

\subsection{GAN objective function}
The proposed loss function includes RaGAN loss \cite{30}, ${\cal L}_{gt}$ loss and ${\cal L}_{fe}$ loss:
\begin{small}
\begin{equation}
{{\cal L}}_{FGAN} = {\cal L}^{RaSGAN}_D + {\cal L}_{G}^{RaSGAN} + {\lambda_{gt}}{{\cal L}}_{gt}(G) + {\lambda_{fe}}{{\cal L}}_{fe}(G)
\end{equation}
\end{small}
\ The RaGAN loss can be expressed as:
\begin{small}
\begin{align}
{\cal L}^{RaSGAN}_D=-{\mathbb{E}_{x_r}}[\log (\tilde{D}(x_r))]-{\mathbb{E}_{x_f}}[\log(1-\tilde{D}(x_f))]\\
{\cal L}^{RaSGAN}_G=-{\mathbb{E}_{x_f }}[\log (\tilde{D}(x_f))]-{\mathbb{E}_{x_r }}[\log(1-\tilde{D}(x_r))]
\end{align}
\end{small}where $\tilde{D}(x_r)=\textrm{sigmoid}(C(x_r)-{\mathbb{E}_{x_f}}C(x_f))$, $\tilde{D}(x_f)=\textrm{sigmoid}(C(x_f)-{\mathbb{E}_{x_r}}C(x_r))$.
$C(x)$ denotes as the non-transformed discriminator output.
The relativistic discriminator aims to predict the probability that a real image $x_r$ is relatively more realistic than a fake one $x_f$ \cite{30}.
 $\lambda _{gt}$ and $\lambda _{fe}$ are the weight of ${\cal L}_{gt}$ loss and ${\cal L}_{fe}$ loss, respectively.

Considering that the proposed network has two inputs, we explore this option by using ${\cal L}_{gt}$ loss and ${\cal L}_{fe}$ loss:
\begin{small}
\begin{align}
{{\cal L}}_{gt}(G)=\mathbb{E}[||x - G(y)|{|_{\rm{1}}}]\\
{{\cal L}}_{fe}(G)=\mathbb{E}[||x_{fe} - G(y)|{|_{\rm{1}}}]
\end{align}
\end{small}where $x_{fe}$ represents the image enhanced by fusion enhance method. $x$ denotes ground truth and $G(y)$ denotes result produced by the generator.

\section{\textbf{Experiments}}
In this section, we first discuss the detail setup of the proposed method and U45 test dataset.
    We then show the performance of the network by comparing it with the other state-of-the-art methods tested on U45 dataset.
Finally, we conduct an ablation study to demonstrate the effect of each component and carry out application tests on low-level and high-level vision tasks to further demonstrate the effectiveness of the proposed method.

\subsection{Setup}

\begin{figure*}[!htb]
\vspace{-4mm}
\centering
\includegraphics[height=212mm,width=180mm,angle=0]{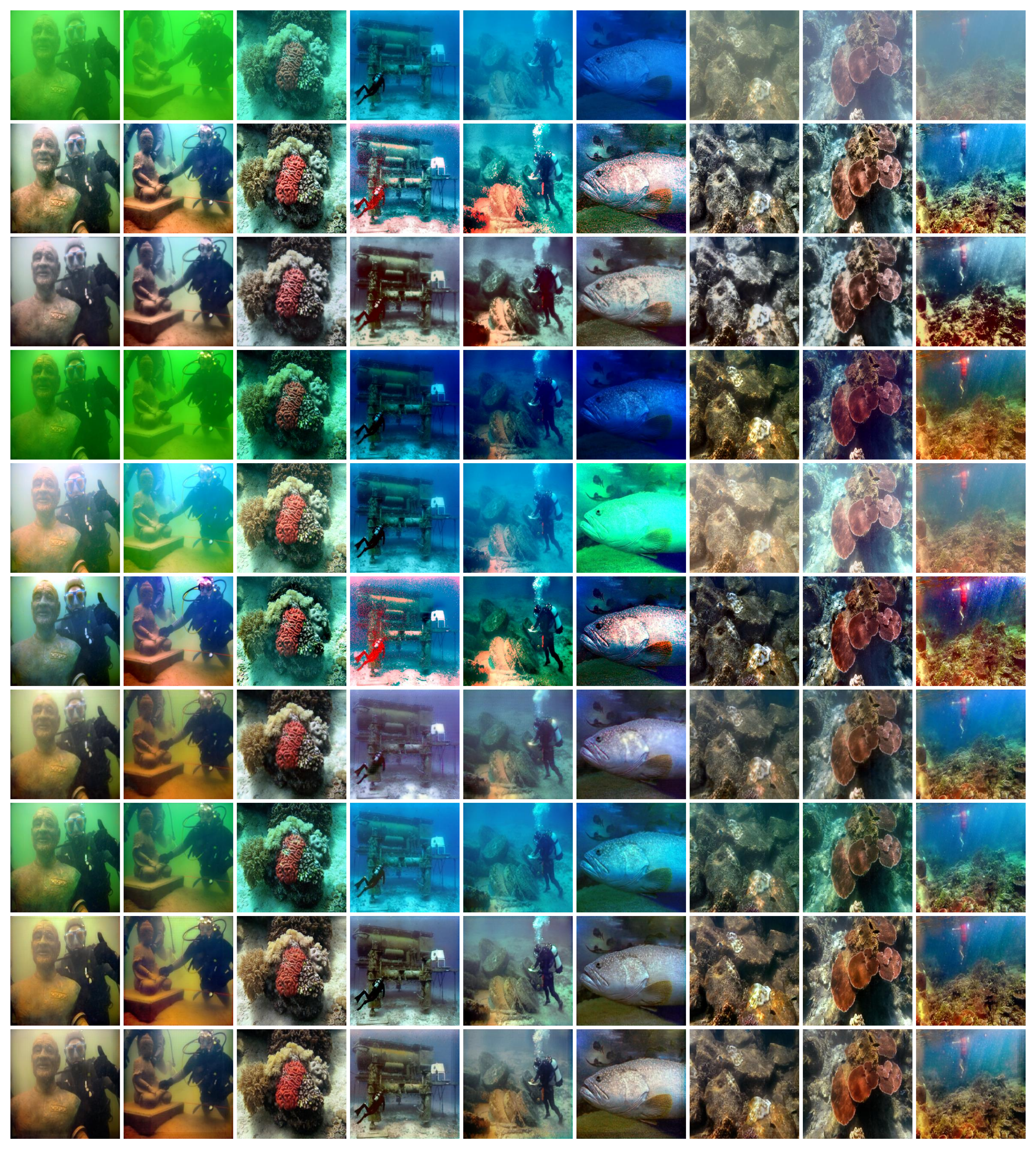}\\
\vspace{-2mm}
\caption{Subjective comparisons on U45. From top to down: Raws, FE, RB, UDCP, UIBLA, DPATN, CycleGAN, WSCT, UGAN, and FGAN. Best viewed with zoom-in.}
\label{fig3}
\end{figure*}

\begin{table*}[htb]
\small
\centering\
 \begin{threeparttable}
 \caption{Testing time and parameters of generator of different enhancement methods.}
  \begin{tabular}{cccccccccc}
  \toprule
     &FE&RB&UDCP&UIBLA&DPATN&CycleGAN&WSCT&UGAN&FGAN\\
  \midrule
   Testing time (s)  &0.1600 &0.2776 &3.6275 &6.2115 &1.1068  &0.1460 &0.1960 &0.0297 &0.0286 \\

  \midrule
   Parameters  &----- &----- &----- &-----  &----- &2.85M &2.85M &38.67M &1.11M  \\		

  \bottomrule
  \end{tabular}
  \label{123}
 \end{threeparttable}
\end{table*}

\subsubsection{Data Set}
Our proposed method is conducted in a paired system by using the 6128 image pairs from the underwater GAN (UGAN) \cite{14},
   where one set of underwater images with no distortion and the another set of underwater images with distortion.
Furthermore, we gather 240 real underwater images from the related papers \cite{3, 4, 5, 9, 14, 18}, Imagenet \cite{31}, SUN \cite{32}, and the sea bed close to the Zhangzi island in the Yellow Sea, China.
    Considering that underwater enhancement task does not have a public and effective test dataset like Set5 \cite{33} or Set14 \cite{34} in image super-resolution task,
thus we carefully select 45 real underwater images named U45 out of the above-mentioned images.
    The U45 is sorted into three subsets of the green, blue, and haze-like categories,
where subsets correspond to the color casts, low contrast and haze-like effects of underwater degradation.
We do not set a low-contrast data set alone because the other three subsets contain various low contrast underwater images.

\subsubsection{Hyperparameter Setting}
In our training process, training and test images have dimensions $256\times256\times3$, $\lambda _{gt}=10$, $\lambda_{fe}=0.5$, Leaky ReLU \cite{35} with a slope of 0.2 and Adam algorithm \cite{36} with the learning rate of 0.0001.
The discriminator updates 5 times per generator update. Batch size is set as 16 due to our limited GPU memory. The entire network was trained on the TensorFlow \cite{37} backend for 60 epochs.

\subsubsection{Compared Methods}
The competitive methods include Fusion Enhance (FE) \cite{4}, Retinex-Based (RB) \cite{5}, UDCP \cite{8}, UIBLA \cite{10}, DPATN \cite{11}, CycleGAN \cite{24}, Weakly Supervised Color Transfer (WSCT) \cite{12}, and UGAN \cite{14}.
For fair comparisons, all the evaluations are implemented on the $256\times256$ image, and the deep learning-based methods are trained on the same training data from the literature \cite{14}.

\subsection{Subjective and objective assessment}

The enhanced results of different methods tested on U45 are shown in Fig. \ref{fig3}.
   The entire results and scores including U45 are available in \url{https://github.com/IPNUISTlegal/underwater-test-dataset-U45-}.

   FE can correct both green and haze-like underwater images well. However, the inaccurate color correction algorithm used by fusion enhance method causes obvious reddish color shift in bluish scenes.
RB fails to deal with image brightness well in most underwater scenes because the same color correction is applied to the three RGB channels, which is inappropriate for underwater scene.
   We notice that UDCP aggravates the blue-green effect but effectively reduces haze effects.
The results of UIBLA are unnatural, especially in the blue scene, because the prior of the background light and the medium transmission estimate is suboptimum.
   The prior-aggregated DPATN is the better choice for haze-like images rather than green and blue images.
For example, blue underwater images enhanced by DPATN introduce unpleasant artifacts and color casts.
   In this paper, CycleGAN and WSCT are trained on image pairs rather than the weakly supervised manner.
WSCT tends to generate inauthentic results in some underwater scenarios because they remain cycle consistency loss more suitable for image to image translation
and abandon the ${\cal L}_{1}$ loss employed for giving results some sense of ground truth, where the WSCT presents visually less appealing than the images appeared in CycleGAN.
   By a general visual inspection it can be noticed that the proposed method is able to correct color distortion and preserve image details than UGAN in green and blue scene.
For example, the coral of green scene presents a bright red, and the fish of blue scene presents visually appealing natural color.
   In addition, the raw underwater video and enhanced video by the proposed method are presented at \url{https://youtu.be/JZTFnBHNGr0}.

In order to evaluate the underwater images, we choose underwater color image quality evaluation (UCIQE) \cite{38} and underwater image quality measure (UIQM) \cite{39}.
   The UCIQE utilizes a linear combination of chroma, saturation and contrast to quantify the nonuniform color cast, blurring, and low contrast, respectively.
The UIQM comprises three properties of underwater images, such as the underwater image colorfulness measure (UICM), the underwater image sharpness measure (UISM), and the underwater image contrast measure (UIConM).
  Higher values of UCIQE and UIQM denote better image quality.

TABLE \ref{tab0} gives the quantitative scores of the compared methods averaged on green underwater images, blue underwater images, haze-like water images, and the entire U45.
   The best result and second best result of UCIQE and UIQM is denoted by the red font and blue font, respectively.
FE yields inconsistent results on UCIQE and UIQM assessments and ranks the low place under UIQM because the inaccurate color correction algorithm causing the decreased value of UICM.
   DPATN aggregated both prior (i.e., domain knowledge) and data (i.e., haze distribution) information boosts UCIM scores of all categories and UCIQE scores of haze-like images, but these results are visually relatively poor.
The proposed method has advantages in correcting color casts and stably obtains high UIQM values in all subsets and U45.
   Simultaneously, FGAN obtains better subjective perception at the cost of the part decreased performance of UCIQE.
Last but not least, the novel dark channel prior loss \cite{13} can remove the distinctive appearance between a hazy image and its clear version in a dark channel.
  However, the hyperparameter tuning still requires much time to improve the performance. Thus, we leave this part in our future work.

We note that there exist discrepancies between the subjective and objective assessment in same underwater scene.
  And, UCIQE and UIQM may yield inconsistent assessments on same dataset, which has been found in the literature \cite{3}.
Recently, Wang et al. \cite{40} proposed an imaging-inspired no-reference underwater color image quality assessment metric, dubbed the CCF,
which shows limitations on balance of human visual perception and objective assessment because CCF obtains the extremely high place in visually poor UDCP results.
   Compared to state-of-the-art opinion-aware blind image quality assessment methods,
   the superior quality-prediction performance of IL-NIQE method \cite{41} is demonstrated to without the need of any distorted sample images nor subjective quality scores for training.
However, it is hard to find high quality underwater images used to learn the pristine MVG model used to create IL-NIQE.
   Besides, researchers also develop deep learning to image quality assessment \cite{42,43,44}, which obtains remarkable performance and could apply to underwater scene in the future work.
In summary, the evaluation of underwater image quality needs more appropriate and effective metric related with human visual perception.

\begin{figure*}[!htb]
\vspace{-6mm}
\centering
\includegraphics[height=52mm,width=70mm,angle=0]{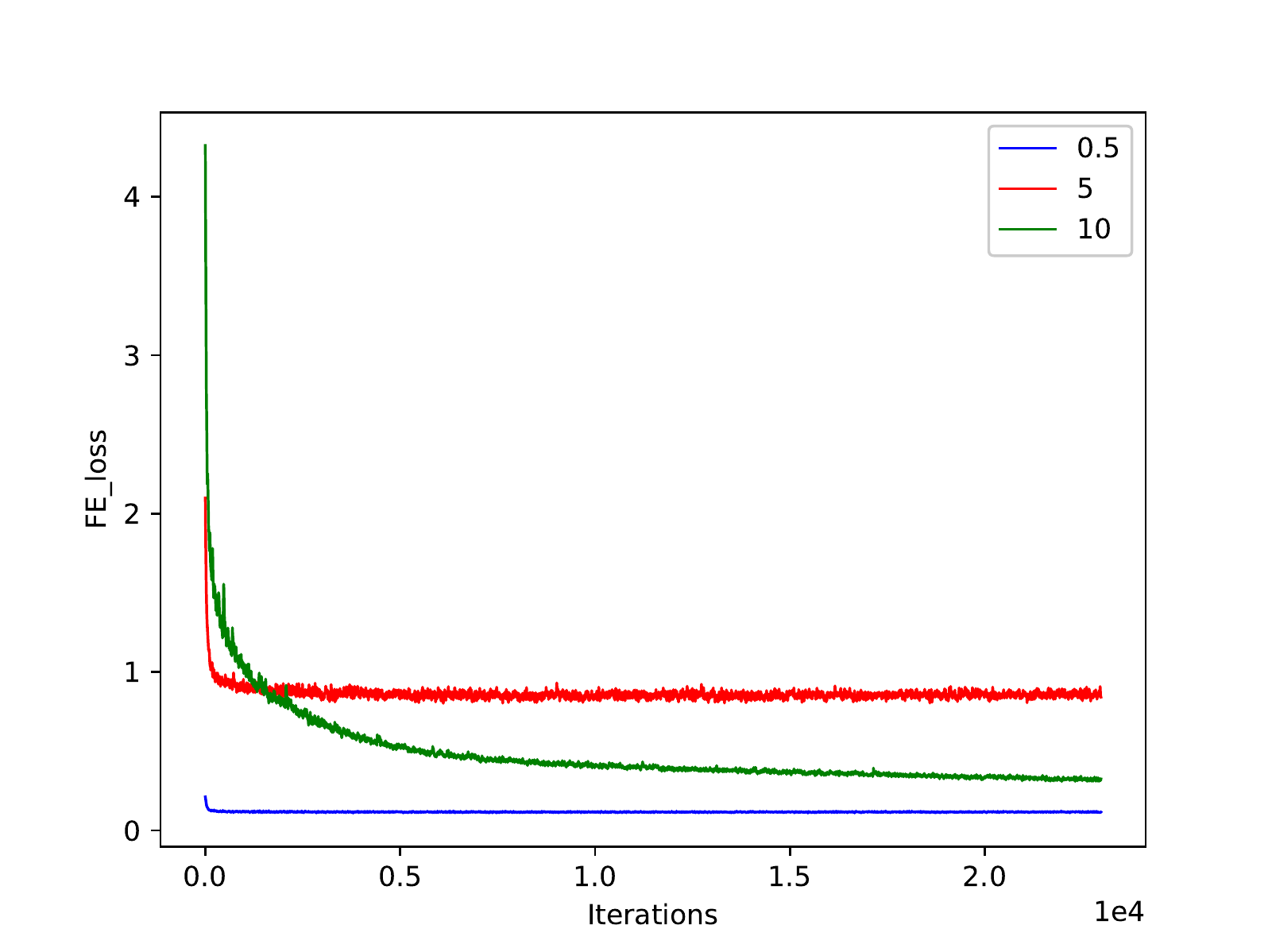}
\includegraphics[height=52mm,width=70mm,angle=0]{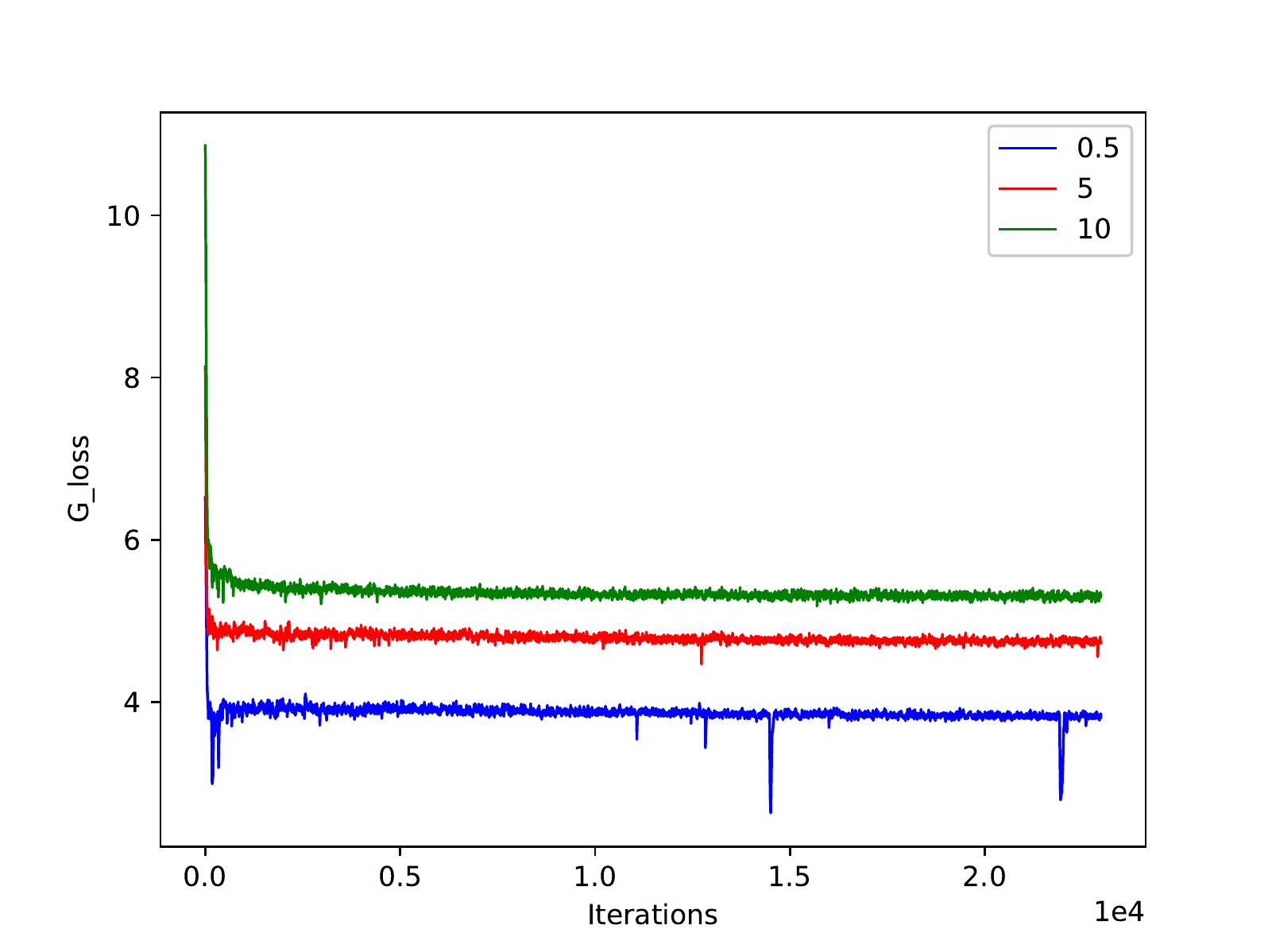}\\
\vspace{-2mm}
\caption{The graph plot of ${\cal L}_{fe}$ loss and total $G$ loss. Blue lines denote $\lambda_{fe}=0.5$. Red lines denote $\lambda_{fe}=5$. Green lines denote $\lambda_{fe}=10$.}
\label{fig4}
\end{figure*}

\begin{figure*}[!htb]
\vspace{-4mm}
\centering
\centering
\subfigure[$\lambda_{fe}$=0]
{\includegraphics[height=1.3in,width=1.3in,angle=0]{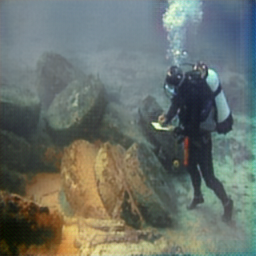}}
\subfigure[$\lambda_{fe}$=0.5]
{\includegraphics[height=1.3in,width=1.3in,angle=0]{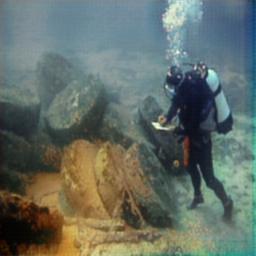}}
\subfigure[$\lambda_{fe}$=5]
{\includegraphics[height=1.3in,width=1.3in,angle=0]{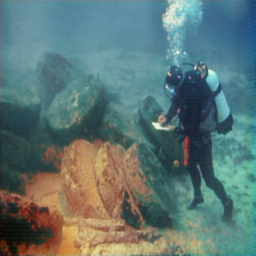}}
\subfigure[$\lambda_{fe}$=10]
{\includegraphics[height=1.3in,width=1.3in,angle=0]{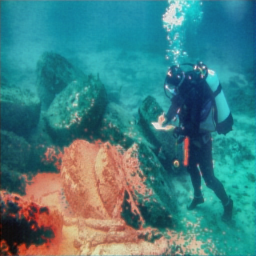}}
\subfigure[FE]
{\includegraphics[height=1.3in,width=1.3in,angle=0]{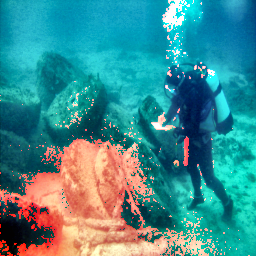}}\\
\vspace{-2mm}
\caption{The enhanced result by different variants and FE.}
\label{fig5}
\end{figure*}

\begin{figure*}[!htb]
\vspace{-4mm}
\centering
\includegraphics[height=52mm,width=70mm,angle=0]{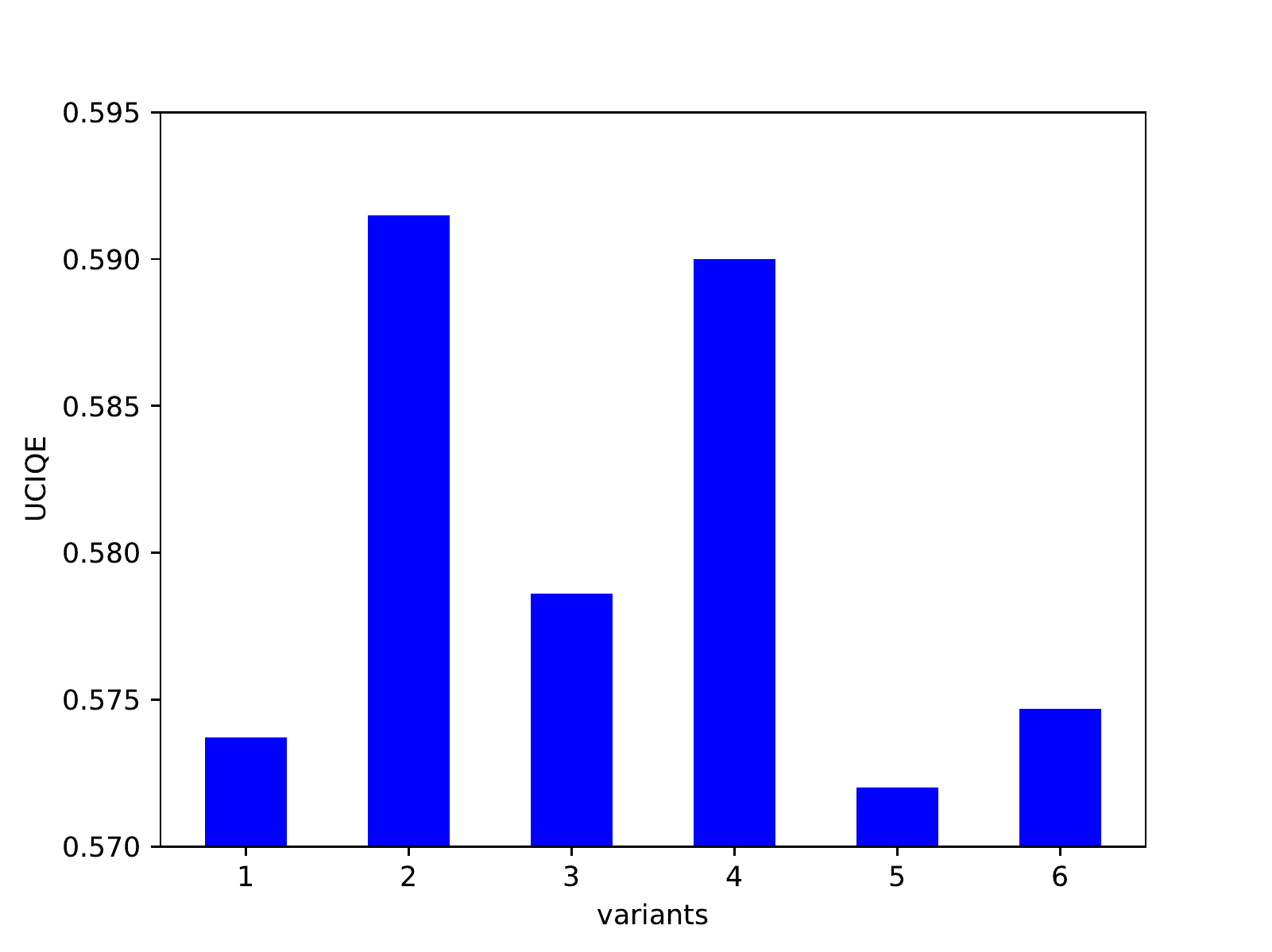}
\includegraphics[height=52mm,width=70mm,angle=0]{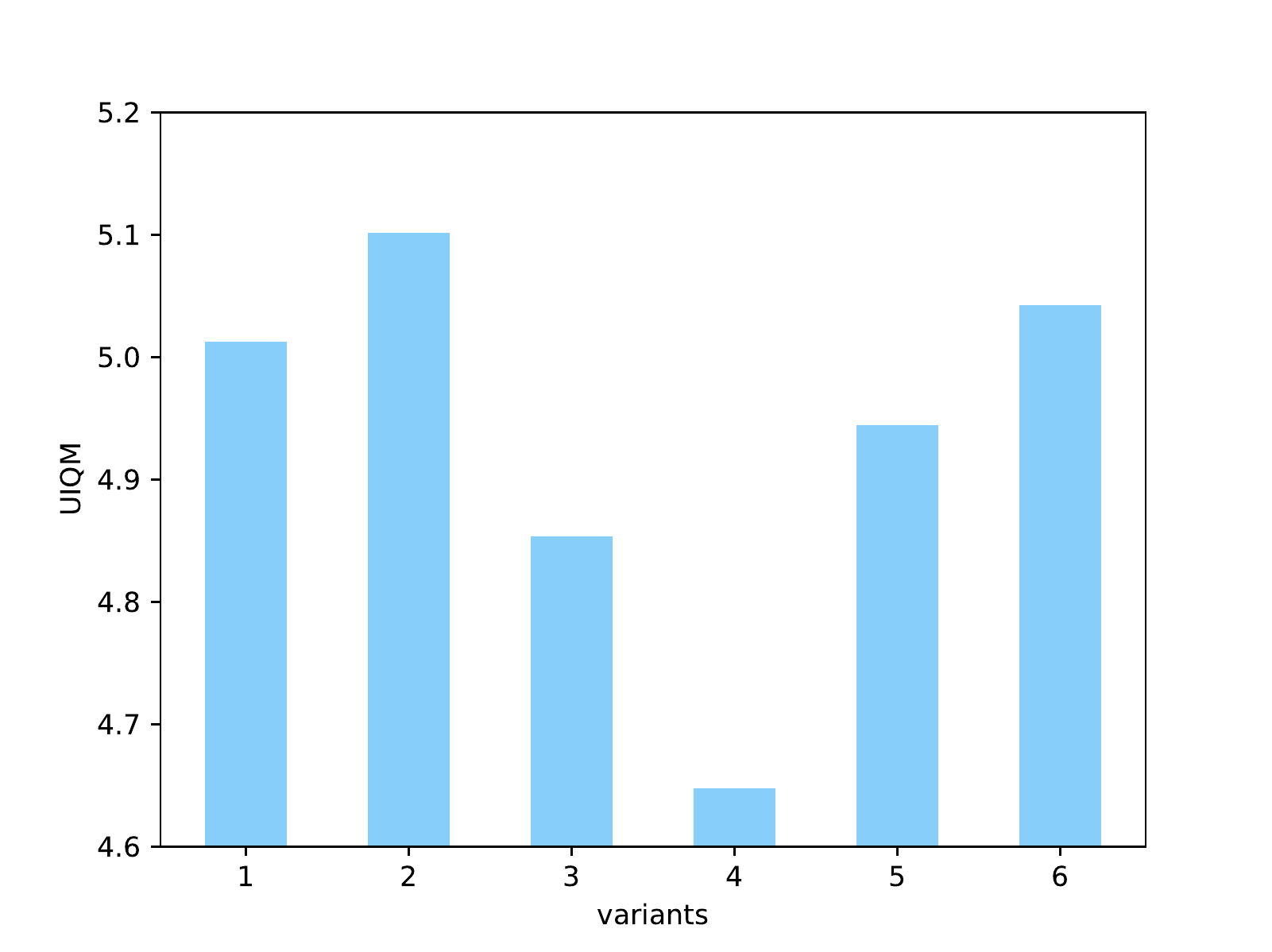}\\
\vspace{-2mm}
\caption{ The graph plot of UCIQE and UIQM. Number 1 to 6 are $\lambda_{fe}$=0, $\lambda_{fe}$=0.5, $\lambda_{fe}$=5, $\lambda_{fe}$=10, -SN and -BB, respectively.}
\label{fig6}
\end{figure*}

\subsection{Testing time and Parameters}
TABLE \ref{123} compares the average testing time of different methods on the same computer with Intel(R) i5-8400 CPU, 16GB RAM.
    Even in the case of combining two inputs, the proposed FGAN owns very fast processing speed with a size of $256\times256$ within 0.0286s.
    Simultaneously, the generator of FGAN contains fewer parameters than other generator of deep learning methods.
UGAN employs many convolution layers with 512 kernels, causing too many network parameters.
Besides, the network of WSCT is based on CycleGAN with 9 blocks for $256\times256$ resolution training images, thus they have the exact same pararmeters.

\begin{figure}[!htb]
\centering
\centering
\includegraphics[height=1in,width=1in,angle=0]{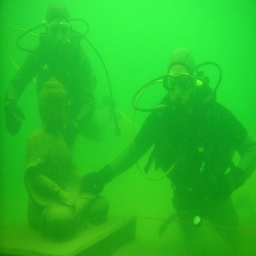}
\includegraphics[height=1in,width=1in,angle=0]{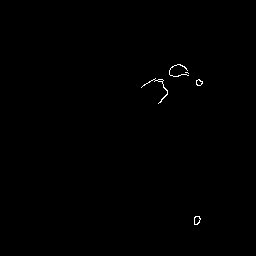}
\includegraphics[height=1in,width=1in,angle=0]{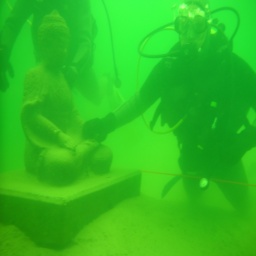}\\
\vspace{1mm}
{\includegraphics[height=1in,width=1in,angle=0]{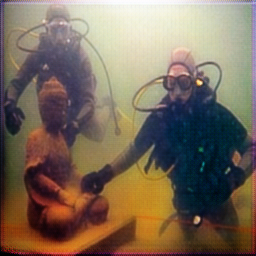}}
{\includegraphics[height=1in,width=1in,angle=0]{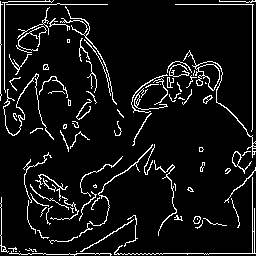}}
{\includegraphics[height=1in,width=1in,angle=0]{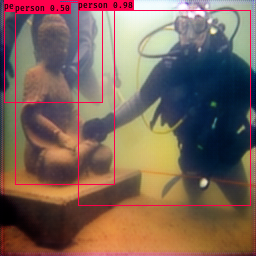}}\\
\caption{Application tests. From left to right and top to down: raw image, canny edge detection of raw image, object detection of raw image, enhanced image, canny edge detection of enhanced image and object detection of enhanced image.}
\label{fig7}
\end{figure}

\subsection{Ablation Study and Application tests}

Ablation study aims to reveal the effect of each component. We carry out the different variants tested on the U45 dataset.
\begin{enumerate}
\item FGAN has different $\lambda_{fe}$ value,
\item FGAN removes spectral normalization (-SN),
\item FGAN replaces all basic blocks with four layers of $3\times3$ kernel size, 256 convolution kernels, and stride 1 (-BB).
\end{enumerate}

As shown in Fig. \ref{fig4}, ${\cal L}_{fe}$ loss has rapid convergence with the increasing numerical index of weight $\lambda_{fe}$.
    Simultaneously, the total $G$ loss decreases with the decreasing numerical index of weight $\lambda_{fe}$.
As depicted in Fig. \ref{fig5}, we notice that when the $\lambda_{fe}$ increases, ${\cal L}_{fe}$ loss plays more important role in training phase and
  the enhanced result is trained to be near the fusion enhance method.
Fig. \ref{fig6} depicts the average UCIQE and UIQM of different variants on U45 test dataset.
For better subjective and objective assessment, the proposed network takes $\lambda_{fe}=0.5$ as the final version.
   Furthermore, we observe that both basic blocks and spectral normalization could improve the performance.

Some application tests, including canny edge detection \cite{45} and object detection, are employed to further demonstrate the effectiveness of the proposed method.
As shown in the Fig. \ref{fig7}, the pre-trained YOLO-V3 model \cite{46} fails to capture objects with a raw image.
        The detection performance of the enhanced images has an obvious improvement, such as sculpture and diver, and more edge detection features are rendered.

\section{\textbf{Conclusion}}
This paper sets up an effective and public U45 test dataset including the color casts, low contrast and haze-like effects of underwater degradation and
presents a fusion adversarial network for underwater image enhancement.
  The proposed network combined with basic blocks and multi-term loss function could correct color effectively and produce visually pleasing enhanced results, which is the first attempt to blend two inputs in generative adversarial network underwater tasks.
Numerous experiments on U45 and an ablation study are conducted to demonstrate the superiority of the proposed method. Besides, the low-level and high-level vision tasks further demonstrate the effectiveness of the proposed method.

\ifCLASSOPTIONcaptionsoff
  \newpage
\fi

\bibliographystyle{IEEEtran}%
\bibliography{reference}

\end{document}